# New design and preliminary tests of a novel matching section for dielectric-loaded accelerating structures

Yelong Wei

[1]*National Synchrotron Radiation Laboratory, University of Science and Technology of China, Hefei, Anhui, 230029 China*

Abstract: In our previous studies on the development of a new type of matching section for dielectric-loaded accelerating (DLA) structures, a significant discrepancy was found between measurements and simulations [*Y. Wei, et al., Phys. Rev. Accel. Beams 25, 041301, 2022*]. This discrepancy was caused by the fabrication errors on the dielectric dimensions. Through introducing the measurement techniques during the fabrication processing on the dielectric tube, we show that this discrepancy can be improved. A new prototype of X-band dielectric tube was then fabricated. The preliminary cold-tests are performed for this dielectric tube without coating.

## I. Introduction

Despite the urgent need for a TeV-class Linear Collider [1-4] in high energy physics, a clear path to realizable and affordable accelerator technologies has yet to be realized. To date, along with laser and beam plasma wakefield accelerators [5-8], another promising concept is the dielectric-loaded accelerating (DLA) structures which utilize dielectrics to slow down the phase velocity of travelling waves in the vacuum channel. A DLA structure comprises a simple geometry where a dielectric tube is surrounded by a conducting cylinder. Because of its simplicity, a DLA structure can be easily scaled from microwave frequency to THz regime.

The DLA structures were initially proposed in the 1940s [9-12], and experimentally demonstrated in the 1950s [13-15]. Since then, disk-loaded metallic structures have prevailed for accelerator research and development because of their high quality factor and high field holding capability. Thanks to remarkable progress in new ceramic materials with high dielectric permittivity ($\varepsilon_r > 20$), low loss ($\tan\delta \leq 10^{-4}$) [16-18], and ultralow-loss ($\tan\delta \leq 10^{-5}$) [19-20], studies on DLA structures are gradually being revived. For example, fused silica, chemical vapor deposition (CVD) diamond, alumina and other ceramics have been studied as materials for DLA structures [21-22] at Argonne National Laboratory [23-25]. In the last two decades, different kinds of DLA structures with improved performance have been reported, such as a hybrid dielectric and iris-loaded accelerating structure [26] reducing the ratio of the peak electric field to the average accelerating field to near unity, a dual-layered dielectric structure [27] and a multilayered dielectric structure [28] achieving a small RF power attenuation, a disk-and-ring tapered accelerating structure [29] enabling a wide range of phase velocities, a dielectric disk accelerating structure [30] and a dielectric assist accelerating structure [31-34] realizing a high RF to beam efficiency.

Since the late 1980s researchers from SLAC and KEK have been pioneering the development of accelerating structures at X-band (11.424 GHz) in the framework of Next Linear Collider (NLC) and Global Linear Collider (GLC) project [35-38]. An accelerating gradient of 65 MV/m with 400 ns long pulses was successfully demonstrated [39-41] at the end of the NLC/GLC program. In 2004, International Technology Review Panel (ITRP) made a key decision to select L-band (1.3 GHz) superconducting technology for the International Linear Collider (ILC) [42]. There was a slowdown for the development of X-band technology afterwards. However, in 2007 CERN decided to lower the frequency of the Compact Linear Collider (CLIC) to 12 GHz (previously at 30 GHz) [2], resulting in a renewed and vigorous interest in X-band technology. In 2010, the undamped design of CLIC X-band travelling-wave structures named T24 has obtained a gradient of 120 MV/m with a pulse length of 200 ns [43-44]. In 2021, a beam-driven X-band travelling-wave structure achieved a gradient of 310 MV/m with a pulse full-width at half-maximum of 6 ns at Argonne National Laboratory [45]. These gradients are far beyond those reached with the present S-band and C-band technology. Today X-band technology is rapidly expanding in the communities of linear collider [46-53], light-source [54-58], medical therapy [59-61].

Building on these developments, a DLA structure operating at X-band appears to be very promising for future linear accelerators. It has been a practical issue for efficiently coupling external RF power into a cylindrical DLA structure with an outer diameter much smaller than the circular waveguide. One scheme separating the RF coupler from the DLA structure with a central accelerating section and two tapered matching sections has been successfully demonstrated in many high-power experimental studies [62-66]. This scheme employs a tapered dielectric matching section with a length of >30 mm, which occupies valuable longitudinal space. Recently a compact, low-field, broadband matching section with a length of around 2 mm was proposed and studied [67-68]. However, a significant discrepancy due to fabrication errors on the dielectric matching section was found between measurements and simulations. Through introducing the measurement techniques during the fabrication processing, we show in this paper this discrepancy can be improved. Section II presents detailed RF design of an X-band DLA structure using $MgTiO_3$ ceramic. Section III shows the preliminary low-power RF measurement for the full-assembly prototypes. Section IV gives the conclusions.

## II. Simulation Studies

In this section, the RF properties of an X-band DLA structure with a central accelerating section and two tapered matching sections (see Fig. 1) are described in detail.

As shown in Fig. 1, the DLA structure has a dielectric tube enclosed by a clamped copper outer jacket. In previous studies [68], an assembly of DLA structure and two $TE_{10}$-$TM_{01}$ mode converters with choke geometry (see Fig. 1) has been tested and a significant discrepancy was found between measurements and simulations. Through detailed analysis, this discrepancy was caused by the fabrication errors on the dielectric tube, especially on the tapered matching sections. A new dielectric tube is then fabricated in order to improve the fabrication errors.

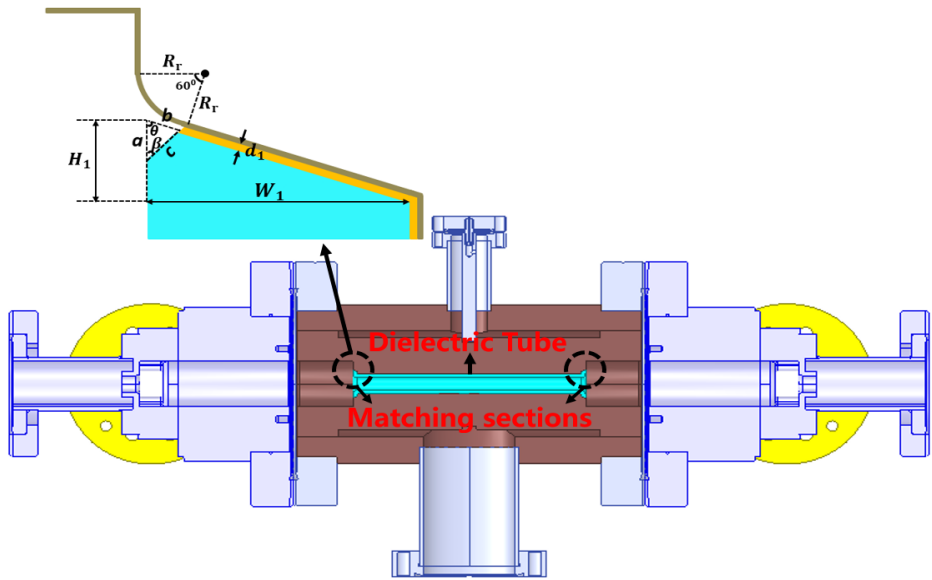

Figure 1. Conceptual illustration of an externally- powered DLA structure and two $TE_{10}$-$TM_{01}$ mode converters with choke geometry

## A. A New DLA Structure

MgTiO$_3$ ceramic, with good thermal conductivity and ultralow power loss, which has been studied in [22], is still chosen as the dielectric material for our new dielectric tube. An accurate measurement of the dielectric properties has to be performed before using such a ceramic for our RF design. Using the TE$_{01\delta}$ silver-plated resonator technique, dielectric constant $\varepsilon_r$ and loss tangent $\tan\delta$ of sample coupons from the dielectric rods as for the fabrication are measured. A dielectric constant $\varepsilon_r = 17.0$ and a loss tangent $\tan\delta = 3.43 \times 10^{-5}$ (having error bars 0.6% of the nominal value) are obtained for the RF design of the new DLA structure.

As the copper jacket and two TE$_{10}$-TM$_{01}$ mode converters with choke geometry from the previous studies [68] are reused for the new dielectric tube, the inner radius is calculated to be $R_{\rm in} = 3.0195$ mm when the outer radius is set to be $R_{\rm out} = 4.6388$ mm for an operating frequency of $f_0 = 11.994$ GHz. The group velocity obtained is $v_g = 0.065c$, where $c$ is speed of light. A quality factor of $Q_0 = 2796$ and a shunt impedance of $R_{\rm shunt} = 25.9$ MΩ/m are also derived for such a new DLA structure using HFSS [69].

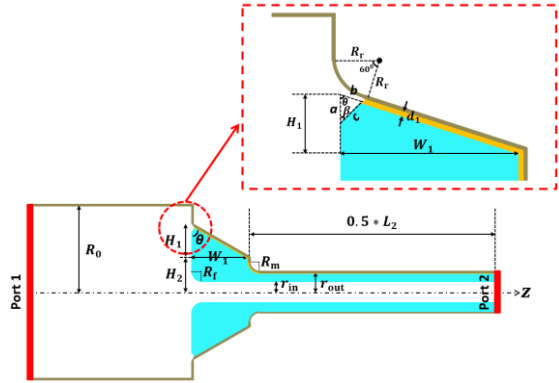

Figure 2. Longitudinal cross section of a circular waveguide, a compact dielectric matching section, and an accelerating section.

Figure 2 shows cross-sectional view of the new DLA structure with a half-length. Through optimization studies, a new dielectric matching section is obtained: $R_{\rm in} = 3.02$ mm, $W_1 = 2.012$ mm, $\theta = 60°$, $\beta = 60°$, $a = b = 0.2$ mm. Other geometrical parameters remain unchanged with the old DLA structure, which can be seen in Table 1.

Table 1. Dimensions for different DLA structures

| Geometrical parameters | Old DLA structure | New DLA structure |
|---|---|---|
| $R_{\rm f}$ [mm] | 2.0 | 2.0 |
| $R_{\rm m}$ [mm] | 0.5 | 0.5 |
| $H_2$ [mm] | 2.74 | 2.74 |
| $W_1$ [mm] | 2.035 | 2.012 |
| $H_1$ [mm] | 1.175 | 1.1616 |
| $\theta$ | 60° | 60° |
| $\beta$ | 45° | 60° |
| $a$ [mm] | 0.254 | 0.2 |
| $b$ [mm] | 0.186 | 0.2 |
| $R_{\rm out}$ [mm] | 4.6388 | 4.6388 |
| $R_{\rm in}$ [mm] | 3.0 | 3.02 |

Figure 3 (a) shows the simulated electric field distribution for the new DLA structure at an input power of 1 W. Figure 3 (b) indicates that the electric fields near that area are much lower than those of the DLA structure. In this case, this dielectric matching section has the potential to survive at a similar high-power level > 5 MW [62-66].

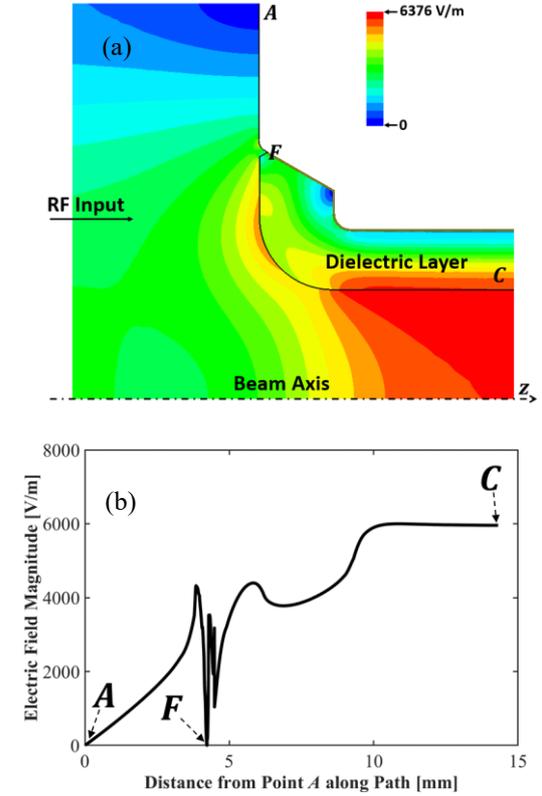

Figure 3. (a) Electric field distribution for the new dielectric matching section. (b) Electric field magnitude along Path $AFC$ which denotes a section of lines and arcs connected by the points $A$, $F$, and $C$, as shown in (a), where the distance of point $A$ is taken as 0 mm.

Figure 4 shows the simulated $S_{11} = -49$ dB and $S_{21} = -0.32$ dB for the new DLA structure at the operating frequency of 11.994 GHz. Using $S_{21} = -0.32$ dB, the coupling coefficient (it is defined by $\eta = 10^{\frac{S_{21}}{5}}$ in this paper) for the whole dielectric structure is calculated to be 86.3%. The $S_{21}$ also has a broad 3 dB bandwidth of more than 1.0 GHz, which allows greater tolerance to potential fabrication errors.

## B. A Full-assembly Structure

A full-assembly structure (see Fig. 5) is modelled by adding the new DLA structure together with the copper jacket and the $TE_{10}$-$TM_{01}$ mode converters with choke geometry. The mode converters with choke geometry have been studied in [67-68]. The RF performance of such a full-assembly structure is described in detail. The whole structure is simulated by analysing the electric field distribution and S-parameters from Port $1'$ to Port $2'$.

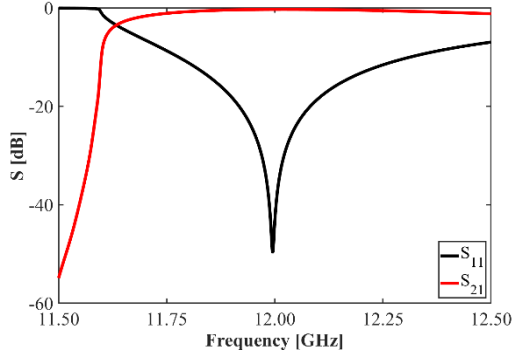

Figure 4. Simulated $S_{11}$ and $S_{21}$ as a function of frequency for the new DLA structure shown in Fig. 2.

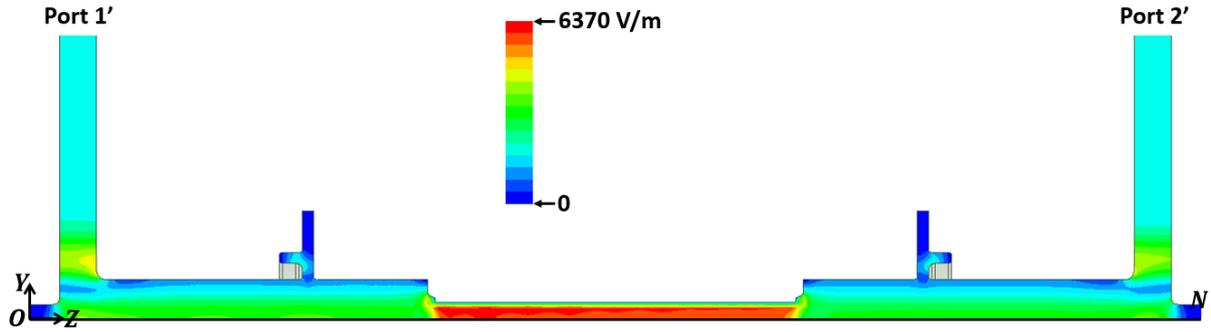

Figure 5. Electric field distribution for the full-assembly structure with a quarter geometry, where line $ON$ is located on the center along $z$-axis.

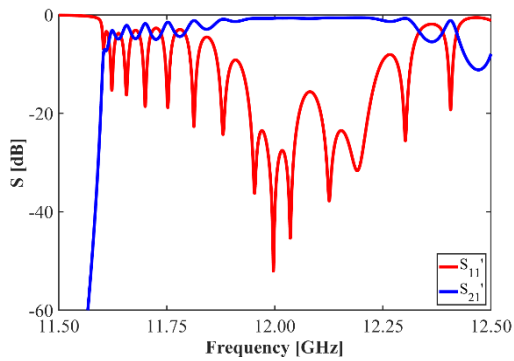

Figure 6. Simulated $S'_{11}$ and $S'_{21}$ as a function of frequency for the full-assembly structure shown in Fig. 5.

Figure 6 gives simulated values of $S'_{11} = -45$ dB and $S'_{21} = -0.71$ dB for the full-assembly structure at the operating frequency of 11.994 GHz. Here $S'_{11}$ and $S'_{21}$ denote the reflection coefficient and transmission coefficient, respectively, between Port $1'$ and Port $2'$, as shown in Fig. 5.

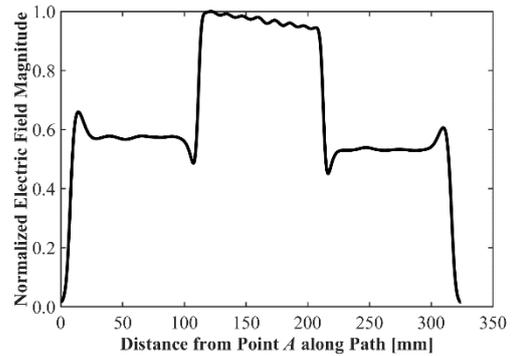

Figure 7. Electric field magnitude along the line *ON* shown in Fig. 5. The distance of point *O* is taken as 0 mm.

Figure 7 shows the electric field magnitude along a line *ON* (see Fig. 5). The electric fields are gradually becoming weaker, due to RF power loss in the dielectric and on metallic surfaces, as the RF fields propagate through the DLA structure. The average accelerating gradient is calculated to be 5781 V/m at an input power of 1.0 W. For an output power of 40 MW from XBOX [70-71] with a pulse width of 1.5 μs and a repetition rate of 50 Hz, an average accelerating gradient of 36.6 MV/m can be achieved for our new DLA structure.

## III. Preliminary Bench Testing

In this section, we described the fabrication processing on the new dielectric tube and the results of low-power RF measurement performed for the full-assembly structure by using a 4-port Vector Network Analyzer (VNA).

A sintered single piece ceramic rod was machined to a dielectric tube with required dimensions in a professional ceramic machining company Insaco [72]. At each step, the measurement techniques were introduced into the fabrication processing on the dielectric tube especially on the width of dielectric matching sections. The new dielectric tube doesn't have a thin coating layer on its outer surface including the central accelerating section and two tapered matching sections, as shown in Fig. 8.

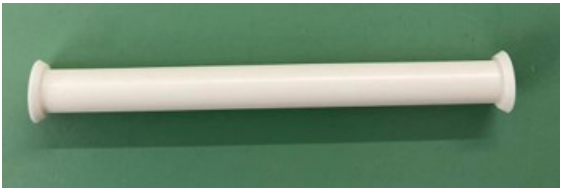

Figure 8. The new dielectric tube doesn't have a coating.

In order to prevent the potential dipole modes and remove the influence of microgap, two power splitters are added into the RF measurement. The 4-port network is then changed to a 2-port network. Figure 9 shows measured $S'_{11} = -5.61$ dB and $S'_{21} = -2.56$ dB at the operating frequency of 11.994 GHz. It is obvious that $S'_{21}$ is much better than one in previous cold-tests [68]. This indicates that the dimensions are improved as compared to previous dielectric tubes. The coating quality is of particular importance for the measurement assembly. Afterwards, the new dielectric tube will be put for coating in the next step. Attention should be paid to even surface roughness from sputtering.

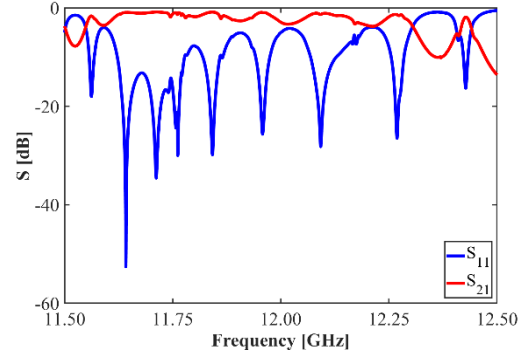

Figure 9. Comparison between simulated and measured S-parameters for the full-assembly structure connected with two power splitters.

## VI. Conclusions

This paper presented a preliminary test of a novel matching section for efficiently coupling the RF power from a circular waveguide to an X-band DLA structure. Through introducing the measurement techniques, a dielectric tube with two tapered matching sections can be fabricated under the tolerances. A prototype of the dielectric tube with the tapered matching sections was mechanically assembled with the mode converters with choke geometry for low-power RF measurement. The measured $S'_{21}$ is much better than one from previous experimental studies. In the next step, the new dielectric tube will be coated with a thin silver layer. More results can be expected in the near future.

## Acknowledgments

The authors would like to thank Dr. Walter Wuensch and Dr. Alexej Grudiev for the useful comments, Dr. Chunguang Jing for the fabrication of dielectric tubes, Dr. Xiaowei Wu for the mechanical and measurement support, and Dr. Mark Ibison for his careful reading of the manuscript.

[72] https://www.insaco.com/